\documentclass[12pt]{iopart}
\usepackage{bbold}
\def\openone{{\mathbb 1}}

\begin{document}

\title{Bounds on action of local quantum channels}
\author{Peter \v{S}telmachovi\v{c}$^{1,2}$ and Vladim\'{\i}r Bu\v{z}ek$^{1,3}$}
\address{
$^{1}$Research Center for Quantum Information, Slovak Academy of Sciences,
 D\'{u}bravsk\'{a} cesta 9, Bratislava, 845 11, Slovakia
 \\
 $^{2}$ {\em Quniverse}, L{\'\i}\v{s}\v{c}ie \'{u}dolie 116, 841 04 Bratislava, Slovakia
\\
 $^{3}$ Abteilung Quantenphysik, Universit\"{a}t Ulm, 89069 Ulm,
Germany
 \\
}

\begin{abstract}
We derive an upper bound on the action of a direct product of two
quantum maps (channels) acting on multi-partite quantum states. We
assume that  the individual channels $\Lambda_j$ affect
single-particle states so, that for an arbitrary input $\rho_j$, 
the distance $D_j (\Lambda_j [ \rho_j ] , \rho_j )$ between the 
input $\rho_j$ and the output $\Lambda_j
[ \rho_j ]$ of the channel is less than $\epsilon$.
Given this assumption  we show that for
an arbitrary {\em separable} two-partite state $\rho_{12}$  the
distance between the input $\rho_{12}$ and the output
$\Lambda_1\otimes\Lambda_2[\rho_{12} ]$ fulfills the bound $D_{12} (
\Lambda_1 \otimes \Lambda_2 [ \rho_{12} ] , \rho_{12} ) \leq
\sqrt{ 2+ 2 \sqrt{(1-1/d_1)(1-1/d_2)}} \, \epsilon$ where 
$d_1$ and $d_2$ are dimensions of first and second quantum 
system respectively. On the contrary, entangled states are transformed in
such a way, that the bound on the action of the local channels is
$D_{12} ( \Lambda_1 \otimes \Lambda_2 [ \rho_{12} ] , \rho_{12} )
\leq 2 \sqrt{2 - 1/d} \: \epsilon$, where $d$ is the dimension of
the smaller of the two quantum systems passing through the channels. Our results
show that the fundamental distinction between the set of separable
and the set of entangled states results into two different bounds
which in turn can be exploited for a discrimination between the two sets of states.
We generalize our results to
multi-partite channels.
\end{abstract}
\maketitle

\section{Introduction}

Investigation of properties of communication channels is more than
ever today a central issue of information sciences. It is generally
accepted that quantum systems have the capacity to carry
information efficiently and any transformation of these systems can be
considered as an action of a quantum channel (see e.g.
Refs.~\cite{stratonovich78, holevo79, holevo98, bennet98, ruskai01}).

Some general questions arising from the transmission of quantum
entaglement through quantum channels has been analyzed by Schumacher in Ref.~\cite{schumacher}.
He has considered a pure entangled state of a pair
of two systems $R$ and $Q$ and the system $Q$ has been subjected to
a dynamical evolution (quantum channel). Schumacher has shown
that the two quantities of interest, the entanglement fidelity $F_e$ 
and the entropy exchange $S_e$, can be related to various other fidelities and entropies
and are connected by an inequality reminiscent
of the Fano inequality of classical information theory.

In this paper we address the question how two local channels
 (each acting independently)
do affect a bi-partite quantum state. This scenario is rather
general and can be applied to a number of situations, e.g. quantum
computation with quantum computer imperfectly isolated from
environment or analysis of quantum error correcting codes. 
In the context of quantum error 
correction this problem  has been
addressed by Knill and Laflamme in Ref.~\cite{Knill} for a
particular case of two qubits and for a particular choice of a distance (fidelity) 
characterizing the change of the bi-partite state. 
In  Ref.~\cite{aharonov} Aharonov {\em et al.} have analyzed errors
for a general model of  quantum computation with mixed states and non-unitary operations.
There however a different measure was introduced. The rationale being that measurable
distinguishability of gates (super-operators) should not increase if we consider additional
quantum systems which do not evolve. Here on the contrary we are not interested in the 
distinguishability of superoperators but rather in the actions of the channels on a given state and 
how to relate these local actions with the change of the global state. 

Specifically, consider a pair of quantum channels 
characterized by maps $\Lambda_1$ and $\Lambda_2$, respectively.
It means that after sending  a quantum system over, for instance, the first channel the final state of
the quantum system (or equivalently the output of
the channel) is $\Lambda_1 [ \rho_1 ]$ where $\rho_1$ is the corresponding input.
Moreover, let the two channels fulfill the following condition
\begin{equation}
\label{con1}
D_j (\Lambda_j [ \rho_j ] , \rho_j ) \leq \epsilon ,  \hspace{0.8cm}
\forall \rho_j \in {\cal S}({\cal H}_j), \;  j=1,2\; ,
\end{equation}
where $D_j(.,.)$ for $j=1,2$ are some distance functions (metric)
defined on the set of all density operators ${\cal S} ( {\cal
H}_1)$ and ${\cal S}({\cal H}_2)$ representing the set of all
physically realizable states of quantum systems passing through
the channels $1$ and $2$, respectively. These conditions restrict
the action of each of the two channels independently of the action
of the other channel. Specifically, the state of a quantum system
affected  by one of the two channels has to be in a small
(epsilon) neighborhood of  the  state describing the quantum
system before the system was sent through the channel. 

The parameter $\epsilon$ quantifies the action of the two
quantum channels. For $\epsilon = 0$ the two channels are
``perfect'' (i.e., noiseless, that is, the information transmitted
via channels is not disturbed) as the output equals to the input
while for $\epsilon$ large the output can be significantly
different from the corresponding input~\footnote{Let us note that
there is no relation between the parameter $\epsilon$ and the
capacity of the channel in general.}. 

The question we would like to address
is, how big the change induced by the two local channels is when
the inputs are correlated. That is let us prepare an arbitrary
initial state $\rho_{12}$ of two quantum systems. The first part
of the jointly prepared  system is sent over the first channel
while the second part is sent over the second
channel. Both channels individually  fulfill the
condition~(\ref{con1}), where, e.g. $\rho_1={\rm Tr}_2\rho_{12}$.
 In what follows we will show that the two-partite action of the channel $\Lambda_1 \otimes \Lambda_2$
for all possible physical states $\rho_{12} \in {\cal S}( {\cal H}_{12})$
fulfills a bound on its action that is determined by single-partite
conditions given by Eq.~(\ref{con1}).

Let us note that the problem can be transformed into the estimation of the map
$\Omega = \Lambda_1 \otimes \Lambda_2 - \openone_{12}$ where the map $\openone_{12}$
is the identity acting on the joint system. If
the distance $D_{12} (.,.)$ as well as distances $D_1(.,.)$ and $D_2(.,.)$ are defined
via a norm  then our task is to estimate the norm $|| \Omega (\rho_{12}) ||$. Similar expressions
for a general class of the so called $p$-norms has been studied extensively for $\Omega$ being a physical map (more specifically
the  product of two physical maps) in Refs.~\cite{amonosov00, king02}.
However, in our case the map $\Omega$  is neither a positive map nor a direct product
of two maps. Due to the fact that the map $\Omega$ is not positive and subsequently not
physical our situation is not applicable to Refs.~\cite{amonosov00, king02} and similar studies.

The paper is organized as follows. In Sec.~2 we introduce necessary definitions and
discuss a particular case of separable states, i.e. the initial state of the joint system
(the system composed of two quantum systems that are sent over the two  quantum channels)
is separable. As a next step we drop any assumptions on the initial
state and analyze the most general case of an arbitrary initial state in Sec.~3.
The results obtained are discussed in Subsection~3.1.
In Sec.~4 we extend our analysis to the case of more than two quantum channels
and illustrate  the nature of changes on a simple example.
Finally, in Sec.~5 we summarize our results and outline possible extensions.

\section{Separable inputs}
\label{ssebsta}

In the formulation of the problem we encounter three different metric (distance) functions:
 $D_1(.,.)$, $D_2(.,.)$ and $D_{12}(.,.)$  acting on different sets and thus measuring
distances between different types of objects. In order to make our discussion explicit
we will consider a specific choice of the distances offered by the norm of the
Hilbert-Schmidt spaces corresponding to the systems
$1$, $2$ and the joint system $12$, respectively~\footnote{The set of all density operators
representing the set of physical states of a quantum system is a subset of a
vector space. In such case it is natural to define the metric (distance function) with the 
help of a norm so that the linear structure of the vector space is respected. 
There are several ways how to introduce a norm on a vector space. However, the
set of all density operators is also a subset of the Hilbert-Schmidt space which is a Hilbert space
and we can use the norm induced with the scalar product of the Hilbert space.
}
\begin{eqnarray}
\label{def1}
D_a(\rho_a, \sigma_a) & \equiv &  || \rho_a - \sigma_a ||_a \nonumber \\
& = &  \sqrt{ {\rm Tr}_a  [ (\rho_a - \sigma_a) (\rho_a - \sigma_a)^\dagger] }  \; .
\end{eqnarray}
The label $a$ denotes the system $1$, $2$ or the joint system $12$ and
$\rho_a, \sigma_a \in {\cal S} ({\cal H}_a)$ are density operators representing possible physical
 states of the system labelled $a$. The norms that we have used to define distances $D_1(.,.)$,
 $D_2(.,.)$ and $D_{12} (.,.)$ are called 2-norms and are only
a particular case of the so called p-norms.
However, due to the fact that we will use only basic properties of the distances $D_1(.,.)$, $D_2(.,.)$
and $D_{12}(.,.)$ we will keep our
derivation as general as possible so that it can be repeated with a broad class of different distances.
 Only in the end we will use the specific choice of distances to derive a tight bound.

Our task is to estimate the distance
\begin{equation}
\label{con0}
D_{12} ( \Lambda_1 \otimes \Lambda_2 [ \rho_{12} ] , \rho_{12} ) \; ,
\end{equation}
for all physically reasonable initial states $\rho_{12} \in {\cal S} ({\cal H}_{12})$ provided
the two maps $\Lambda_1$ and $\Lambda_2$ fulfill the condition~(\ref{con1}). First, note
that for any distance inequality (this follows from the triangle property of a distance) holds
\begin{eqnarray}
\label{exp8}
D_{12} ( \Lambda_1 \otimes \Lambda_2 [ \rho_{12} ] , \rho_{12} ) &  \leq &
D_{12} ( \Lambda_1 \otimes \openone [ \rho_{12} ] , \rho_{12} )  \nonumber \\
& + &  D_{12} ( \openone \otimes \Lambda_2 [ \Lambda_1 \otimes \openone [ \rho_{12} ] ] .
 \Lambda_1 \otimes \openone [\rho_{12} ] ) \,  .
\end{eqnarray}
It means that instead of considering the case with two local channels it is sufficient to consider
only an action of a single local channel acting on one of the two subsystems and estimate the distance
$
D_{12} ( \Lambda_1 \otimes \openone [ \rho_{12} ] , \rho_{12} )
$.

We start with the simplest case - the case of factorizable states of the
form $\rho_{12} = \rho_1 \otimes \rho_2$.
This corresponds to the situation as if the two channels were considered separately
so that the two quantum systems that are  sent through the channels are prepared individually. In this case we exploit the following property
\begin{eqnarray}
\label{j1}
D_{12} (\rho_1' \otimes \rho_2, \rho_1 \otimes \rho_2 ) 
\leq D_1 (\rho_1', \rho_1) \; ,
\end{eqnarray}
of the distances $D_1(.,.)$, $D_2(.,.)$ and $D_{12}$ where  
the operators $\rho_1$, $\rho_1'$ and $\rho_2$  
are density operators representing states
of the first and the second system, respectively.
Let us note that this relation holds even if the distances are defined with any p-norm or even fidelity. Using Eq.~(\ref{j1}) we have that
$ D_{12} ( \Lambda_1 \otimes \openone  [ \rho_{1} \otimes \rho_2  ], \rho_{1} \otimes \rho_2 )
 \leq  D_1 (\Lambda_1 [ \rho_1], \rho_1 ) \; $ and consequently, 
for the initial state of the form $\rho_1 \otimes \rho_2$, the distance
 (\ref{con0}) is always less or at most equal to $2 \epsilon$
\begin{equation}
D_{12} ( \Lambda_1 \otimes \Lambda_2 [ \rho_1 \otimes \rho_2  ] , \rho_1 \otimes \rho_2 )
\leq 2 \epsilon \; ,
\end{equation}
due to Eqs.~(\ref{exp8}) and (\ref{con1}).

The same holds for the initial state $\rho_{12}$ of the form
$\rho_{12} = \sum_i \alpha_i \rho_1^i \otimes \rho_2^i$, where $\alpha_i \geq 0$,
$\sum_i \alpha_i = 1$ and $\rho_1^i$ and $\rho_2^i$ denote density operators
of the system $1$ and $2$ respectively, that follows from the linearity of the map
 $\Lambda_1 \otimes \Lambda_2$ 
\begin{eqnarray*}
D_{12} (\Lambda_1\otimes \Lambda_2 && \hspace{-0.5cm}
[ \sum_i \alpha_i \rho_1^i \otimes \rho_2^i ] 
 , \sum_i \alpha_i \rho_1^i \otimes \rho_2^i) \\
& = &  D_{12} (\sum_i \alpha_i \Lambda_1\otimes \Lambda_2 
[ \rho_1^i \otimes \rho_2^i ] , \sum_i \alpha_i \rho_1^i \otimes \rho_2^i) \; , 
\end{eqnarray*}
and the fact that the distance $D_{12}(.,.)$ is {\em jointly convex}
, that is 
\begin{eqnarray*}
D_{12} (\sum_j \alpha_j \Lambda_1\otimes \Lambda_2 && \hspace{-0.5cm}
[ \rho_1^j \otimes \rho_2^j ] ,
\sum_j \alpha_j \rho_1^j \otimes \rho_2^j) \\
& \leq & \sum_j \alpha_j D_{12} ( \Lambda_1 \otimes \Lambda_2 
[ \rho_1^j \otimes \rho_2^j ] , \rho_1^j \otimes \rho_2^j) \; .
\end{eqnarray*}
The last expression is a sum of terms where each term is bounded 
by $2 \epsilon$ and the sum of the coefficients $\alpha_i$ 
is equal to unity. In consequence we obtain the bound
\begin{eqnarray}
\label{exp10}
D_{12}(\Lambda_1 \otimes \Lambda_2 [ \sum_i \alpha_i \rho_1^i \otimes \rho_2^i ]
, \sum_i \alpha_i \rho_1^i \otimes \rho_2^i) \leq 2 \epsilon \; ,
\end{eqnarray}
for an arbitrary separable state.

\subsection{Hilbert-Schmidt distance}

The bound on the action of a product of two quantum channels on separable states (\ref{exp10}) 
is valid for any triple of distances 
$D_1(.,.)$, $D_2(.,.)$ and $D_{12}(.,.)$ that 
satisfy relation~(\ref{j1}) (the distance  $D_2(.,.)$ has to fulfill the relation~(\ref{j1}) 
with swapped labels $1$ and $2$) and in addition 
the distance $D_{12}(.,.)$ has to be {\em jointly convex}.
That is, the bound is valid if $D_1 (.,.)$, $D_2(.,.)$ and $D_{12} (.,.)$ 
are trace distances~\footnote{The 
trace distance is defined with the help of the 1-norm and $D(\rho, \sigma)$ is equal to 
the sum of eigenvalues of the positive operator
$| \rho -\sigma|$ where $| \rho - \sigma|  \equiv \sqrt{(\rho - \sigma)^\dagger(\rho -\sigma)}$ and 
$\rho, \sigma \in {\cal B}({\cal H})$.} 
or, more generally, the distances defined with p-norms or even fidelity.
The question is whether it is possible to derive a better (tighter) bound 
or, in other words, whether the bound  is optimal. 
For the trace distances the bound is optimal indeed and it can be shown that there is 
a pair of maps such that the bound is saturated. In what follows 
we will show that for the distances introduced in Eq.~(\ref{def1})  
the bound can be further optimized.

Let $\rho_1 = 1/d_1 \openone +  \bar{c} . \bar{\sigma}$ 
be an input of the  channel $1$.  We have expressed the state of the system
labeled as $1$ using the identity operator $\openone$ and $d_1^2 -1$ generators 
$\bar{\sigma} = \{ \sigma_1, \sigma_2, \ldots \}$  of the group $SU(d_1)$ 
multiplied with the complex unity  
where $d_1$ is the dimension of the Hilbert 
space of the system $1$ and the vector $\bar{c} = \{ c_1, \ldots \}$ 
is a real vector with $d_1^2 -1$ elements. 
In addition we require that the set of operators $\{ \sigma_\alpha \}$ satisfy 
the ortho-normalization condition ${\rm Tr} \, \sigma_\alpha \sigma_\beta = \delta_{\alpha \beta}$. 
After the quantum system has been sent through the quantum  channel $\Lambda_1$ the state 
of the system (the output) can be expressed using the same notation 
$\Lambda_1 [ \rho_1 ] = 1/d \openone + \bar{c}'.\bar{\sigma}$
with new coefficients $\bar{c}'$ where the prime  indicates the fact that the state 
 has been sent through the quantum channel.
Equivalently, $\rho_2 = 1/d_2 \openone + \bar{d}. \bar{\tau}$
is the most general state of the system $2$ where $\bar{\tau}$ are generators of 
$SU(d_2)$ multiplied with complex unity, $d_2$ is the dimension of the Hilbert space
of the system $2$ and the operators $\{ \tau_\beta \}$ satisfy relation 
${\rm Tr} \, \tau_{\beta} \tau_{\omega} = \delta_{\beta \omega}$.

We estimate the distance (\ref{con0}) for an arbitrary separable state and 
the particular choice of distances (\ref{def1}).
Due to the {\em joint convexity} of the distance $D_{12} (.,.)$  and the linearity of the map 
$\Lambda_1 \otimes \Lambda_2$ it is sufficient to consider the  case where the 
state $\rho_{12}$ is a pure state (for more details see the end of the previous section)  
\begin{equation}
\label{eein}
\rho_{12} = ( 1/d_1 \openone + \bar{c} . \bar{\sigma})  \otimes ( 1/d_2 \openone
+ \bar{b}. \bar{\tau} )
\end{equation}
where $\bar{c}. \bar{c} = (1-1/d_1)$ and $\bar{b} .\bar{b} = (1-1/d_2)$. 

In this case we do not use the relation Eq.~(\ref{exp8}) which means that 
the two channels are not considered separately and the output of the product of
the two channels $\Lambda_1$ and $\Lambda_2 $ is
\begin{equation}
\label{eeout}
\Lambda_1 \otimes \Lambda_2 [\rho_{12} ]  = 
( 1/d_1 \openone  + \bar{c}' . \bar{\sigma} ) \otimes 
( 1/d_2 \openone + \bar{b}' . \bar{\tau} ) \; .
\end{equation}
Inserting the two expressions, input (\ref{eein}) and output (\ref{eeout}), into the definition 
of the distance~(\ref{def1}) we obtain that
\begin{eqnarray}
D_{12} (\Lambda_1 \otimes \Lambda_2 [ \rho_{12} ], \rho_{12} )  = \nonumber \\  
|| (\bar{c}' -  \bar{c}) . \bar{\sigma} \otimes 1/d_2 \openone
+ 1/d_1 \openone \otimes (\bar{b}' - \bar{b}) . \bar{\tau}
+ \bar{c}' . \bar{\sigma} \otimes \bar{b}' . \bar{\tau}
- \bar{c} . \bar{\sigma} \otimes \bar{b} . \bar{\tau} ||_{12}  \; .
\end{eqnarray}
Last expression squared can be bounded from above by a sum of three terms
\begin{eqnarray*}
|| (\bar{c}' -  \bar{c}) . \bar{\sigma} \otimes 1/d_2 \openone ||_{12}^2
\; + \; || 1/d_1 \openone \otimes (\bar{b}' - \bar{b}) . \bar{\tau} ||_{12}^2 \\
\; + \; [ \;  || (\bar{c}' - \bar{c}) . \bar{\sigma} \otimes \bar{b}' . \bar{\tau} ||_{12}
+|| \bar{c} . \bar{\sigma} \otimes ( \bar{b}' - \bar{b}) . \bar{\tau} ||_{12} \;  ]^2 \; .
\end{eqnarray*}
Observing that  $|| (\bar{c} - \bar{c}' ) . \bar{\sigma} \otimes 1/d_2 \openone ||_{12}^2 =
 1/d_2 \, D^2_1(\Lambda_1 [ \rho_1 ] , \rho_1 )$, 
and equivalently $|| 1/d_1 \openone \otimes (\bar{b} - \bar{b}') . \bar{\tau} ||_{12}^2 = 
1/d_1 D_2 (\Lambda_2 [ \rho_2 ] , \rho_2 )$ and $\bar{b}' . \bar{b}' \leq (1-1/d_2)$ 
we can bound the distance squared with the expression
$1/d_2 \epsilon^2 + 1/d_1 \epsilon^2 + ( \sqrt{1-1/d_1} + \sqrt{1-1/d_2})^2 \epsilon^2$.
Finally, the distance between the input and the corresponding output of the product of the 
two channels fulfills the bound
\begin{equation}
\label{ee14}
D_{12} (\Lambda_1 \otimes \Lambda_2 [ \rho_{12} ] , \rho_{12}) \leq
\sqrt{ 2+ 2 \sqrt{(1-1/d_1)(1-1/d_2)}} \; \epsilon \; ,
\end{equation}
where $d_1$ and $d_2$ are the dimensions of the Hilbert spaces corresponding to
 the quantum systems sent through the channels $1$ and $2$, respectively. Even though 
we have proved the bound for pure separable 
states, we note that the result is valid for an arbitrary separable state due to the linearity 
of the map $\Lambda_1 \otimes \Lambda_2$ and the {\em joint convexity} of the distance
$D_{12} (.,.)$. The bound (\ref{ee14}) is undoubtedly better than the bound (\ref{exp10}) as it has been
derived for a specific choice of distances. In addition,  it can be shown that the bound is 
optimal in the sense that there is a pair of maps $\Lambda_1$ 
and $\Lambda_2$ and a separable state $\rho_{12}$ such that the bound (\ref{ee14}) is saturated
(optimality is discussed in a more detail in Sec.~\ref{conclusion}).

\section{Entangled states}
\label{entstates}

We have seen that if the initial state of the joint system $12$ is factorizable or even separable then
the action of the two channels is bounded by the expression $\sqrt{ 2+ 2 \sqrt{(1-1/d_1)(1-1/d_2)}} \; \epsilon$
It may be tempting to say that the same holds
for an arbitrary state. However, as the next example illustrates, if the joint state of the
two systems $1$ and $2$ is entangled then for certain maps the separable bound can be broken.

Let us consider the Hilbert spaces ${\cal H}_1$ and ${\cal H}_2$ corresponding to the systems $1$ and $2$ to
be two-dimensional spaces. This is the simplest possible case though the physical representations
of such systems are numerous. As an example we
can mention spin one-half particles, polarized photons or particular internal degrees of freedom of an ion.
Let us note that  in quantum
information theory such systems are denoted as qubits since they represent quantum analogue of a classical bit of
information.

Then, any physical state of the system $1$ (or equivalently of the system $2$)
can be written as
$
\rho_1 = \frac{1}{2} (  \openone + \vec{\alpha} . \vec{\sigma} ) 
$,
where $\vec{ \alpha} = (\alpha_1, \alpha_2, \alpha_3) $ is a vector in a three dimensional real vector space
and the three matrices  $\vec{\sigma} = (\sigma_1, \sigma_2, \sigma_3 )$ are the well known Pauli operators.
For the matrix $\rho_1$ to represent  a physical state the norm of the real vector $\vec{\alpha}$
has to be less or equal to 1. It follows that the set of all physically realizable states of the
system $1$ corresponds to  a unit ball (Bloch sphere) in the three dimensional vector space ${\mathbb R}^3$.

The map $\Lambda_1$ we will consider in this particular example is a simple contraction of the ball
representing the set of states such that
\begin{eqnarray}
\label{pl1}
\Lambda_1 : \rho_1 \rightarrow \frac{1}{2} ( \openone + (1-k) \vec{\alpha} . \vec{\sigma} ) \; ,
\end{eqnarray}
where $(1-k)$ is a parameter of the contraction. Physically, the map $\Lambda_1$ 
describes a channel with uncolored (``white'') noise 
since each input state is mixed with the absolute mixture $1/2 \; \openone$ which is the fixed point
of the $\Lambda_1$.
In order to preserve the condition~(\ref{con1}) the parameter
$k$ has to fulfill the relation $k \leq \sqrt{2} \epsilon$. In what follows we assume $k = \sqrt{2} \epsilon$.

In the same way the most general state of the system $2$ is 
$
\rho_2 = \frac{1}{2} ( \openone + \vec{\beta} . \vec{\sigma} )
$, 
where $\vec{\beta} = ( \beta_1, \beta_2, \beta_3)$ is a real vector and $ | \beta | \leq 1$.
The map $\Lambda_2$ has been chosen to be the same as the map $\Lambda_1$
\begin{eqnarray}
\label{pl2}
\Lambda_2 : \rho_2 \rightarrow \frac{1}{2} ( \openone + (1-k') \vec{\beta} . \vec{\sigma} ) \; ,
\end{eqnarray}
with the same contraction parameter $k' = k = \sqrt{2} \epsilon $ so that the condition~(\ref{con1})
is fulfilled in this case too.

To show that the separable bound can be broken we have to  consider an entangled state.
However, we will not consider an arbitrary state
but a very specific one - a  maximally entangled state known as the Bell state of the form
$
\rho_{12} = 1/2 ( | 01 \rangle - | 1 0 \rangle ) ( \langle 01 | - \langle 10 | )
$,
where $0$ and $1$ denote two basis vectors of ${\cal H}_1$ (or ${\cal H}_2$).
For subsequent calculations it is useful to rewrite the state using the Pauli operators
$ \rho_{12} = 1/4 \; ( \openone \otimes \openone - \sigma_1 \otimes \sigma_1 - \sigma_2 \otimes
\sigma_2 - \sigma_3 \otimes \sigma_3 )$. Inserting $\rho_{12}$  into  
Eq.~(\ref{con0}) and using the linearity of
the transformation $\Lambda_1 \otimes \Lambda_2$ as well as  Eq.~(\ref{def1}) the
distance in Eq.~(\ref{con0}) reads
\begin{eqnarray*}
D_{12} ( \Lambda_1 \! \otimes \! \Lambda_2 [\rho_{12} ] , \rho_{12} ) = 
|| \frac{ - k \! - \! k ' \! + \! k k' }{4} \;
\{  \sigma_1 \! \otimes \! \sigma_1 + \sigma_2 \! \otimes \! \sigma_2 + \sigma_3 \! 
\otimes \! \sigma_3 \}  ||_{12} \; .
\end{eqnarray*}
Both constants, $k$ as well as $k'$ are equal to $\sqrt{2} \epsilon$. Neglecting terms
of the order $\epsilon^2$ and evaluating the norm using the scalar product we find
\begin{equation}
D_{12} ( \Lambda_1 \otimes \Lambda_2 [\rho_{12} ] , \rho_{12} ) \approx  \sqrt{6} \; \epsilon \; .
\end{equation}
This result clearly shows that even though the two maps $\Lambda_1$ and $\Lambda_2$ fulfill
 the relations~(\ref{con1}) the map $\Lambda_1 \otimes \Lambda_2$ constructed
as a {\em direct} product of the two maps can affect the states it acts on in a much stronger way.
How much the joint (and particularly entangled) states can change by two arbitrary maps
$\Lambda_1$ and $\Lambda_2$ is  addressed in the next  paragraph.
\vskip 0.5cm

\noindent
Let $\rho_{12}$ be an arbitrary mixed state. The deviation of the output of the channel
$\Lambda_1 \otimes \Lambda_2$ from the input $\rho_{12}$ is characterized
by the distance Eq.~(\ref{con0}). In order to estimate the distance we exploit (as in the case
of separable states) the bound given by Eq.~(\ref{exp8})
\begin{eqnarray}
\label{exp6}
D_{12} ( \Lambda_1 \! \otimes \! \Lambda_2 [ \rho_{12} ] , \rho_{12} ]  \leq
D_{12} (\Lambda_1 \! \otimes \! \openone
[ \rho_{12} ] , \rho_{12} ) 
& + & D_{12} ( \openone \! \otimes \! \Lambda_2 [ \tilde{\rho}_{12} ] , \tilde{\rho}_{12} ) \; ,
\end{eqnarray}
where $\tilde{\rho}_{12} = \Lambda_1 \otimes \openone [ \rho_{12} ]$.
 As we do not make any assumptions
 about neither  the maps $\Lambda_1$ and $ \Lambda_2$ nor the initial  state $\rho_{12}$
 the two states $\tilde{\rho}_{12}$ and $\rho_{12}$ can be  arbitrary physical states
of the joint quantum system, i.e.  arbitrary density operators. It means that taking,
for instance, the first term on the right-hand side of Eq.~(\ref{exp6}) we need to estimate
this term for all possible maps $\Lambda_1$ and all possible states $\rho_{12}$.
This fact allows us to rewrite the bound for (\ref{con0}) in a different way
\begin{eqnarray*}
D_{12} (\Lambda_1 \otimes \Lambda_2 [ \rho_{12} ] , \rho_{12} ) \; \leq 
2 \; {\rm sup}_{\{ \Lambda_1, \rho_{12} \} } || \Lambda_1 \otimes \openone  [ \rho_{12} ] - \rho_{12}  ||_{12} \; ,
\end{eqnarray*}
where the factor $2$ appears because we have two terms in Eq.~(\ref{exp6}) and the
supremum runs over all possible maps $\Lambda_1$ and all initial states $\rho_{12}$~\footnote{
Given the fact that  we have not specified the dimension of neither the system $1$ 
nor the system $2$, the two systems can be different. 
Therefore we should find supremum over all $\Lambda_1$'s and all $\rho_{12}$ of the first
expression in Eq.~(\ref{exp6}) and all $\Lambda_2$'s and all $\tilde{\rho}_{12}$ of the second expression
in Eq.~(\ref{exp6}). However, the results are the same in both cases.}.

A mixed state $\rho_{12}$ can be decomposed into a mixture of pure states
$\rho_{12} =  \sum_k \alpha_k | \psi_k \rangle \langle \psi_k |$.
Using a basic property of the norm (or {\em joint convexity} of the distance)
and the normalization condition $\sum_k \alpha_k = 1$ we can simplify the last expression
and instead of searching for supremum over all possible states $\rho_{12}$ of the joint system ${12}$
it is sufficient to consider pure states only. It means that
\begin{eqnarray}
\label{o21}
D_{12} (\Lambda_1 \! \otimes \! \Lambda_2 [ \rho_{12} ] , \rho_{12} ) \; \leq 
2 \; {\rm sup}_{\{ \Lambda_1, |\psi \rangle \langle \psi | \} }
|| \, \Lambda_1 \! \otimes \! \openone  [  | \psi \rangle \langle \psi |  ]  - 
| \psi \rangle \langle \psi |   \, ||_{12}  ,
\end{eqnarray}
where the supremum runs over all possible maps $\Lambda_1$ and all possible pure states
$| \psi \rangle \langle \psi | \in {\cal S} ({\cal H}_{12})$ of the joint system $12$.
Since we have used only a basic property of the norm the last relation is valid
for any distance defined with the help of a norm (or more generally any distance that is {\em jointly convex}). 
However, in what follows we will use
specific properties of the Hilbert-Schmidt norm and further  results are valid for
that particular choice of the norm only.

Any
pure state $| \psi \rangle \in {\cal H}_{12}$ can be expressed using the Schmidt
basis
\begin{eqnarray}
\label{exp12}
| \psi \rangle = \sum_{k=1}^{n_\psi}  \beta_k | k \rangle_1 \otimes | k \rangle_2 \; ,
\end{eqnarray}
where $\{ | k \rangle_1 \}$ and $\{ | k \rangle_2 \}$ are two sets of orthonormal vectors of
${\cal H}_1$ and ${\cal H}_2$, respectively, and  $\beta_k$ are real positive coefficients.
The integer  $n_\psi$ denotes the number of elements in
the Schmidt decomposition of the given  pure state and is always less or equal
to the dimension of the smaller of the two
Hilbert spaces ${\cal H}_1$ and ${\cal H}_2$. In this particular basis the state
 $\rho_{12} = | \psi \rangle \langle \psi | $ has the  form
\begin{eqnarray}
\label{exp5}
| \psi \rangle \langle \psi | = \sum_{k,l =1}^{n_\psi}  \beta_k \beta_l | k \rangle_1 \langle l |
\otimes | k \rangle_2 \langle l | \; .
\end{eqnarray}
Let us now estimate  the expression $|| \; \Lambda_1 \otimes \openone [ | \psi \rangle \langle \psi | ]
\; - \; | \psi \rangle \langle \psi | \; ||_{12}^2 $ from Eq.~(\ref{o21}). 
Using  Eq.~(\ref{exp5}) for the density operator
 $| \psi \rangle \langle \psi |$ and tracing over the degrees of freedom belonging to
 the second system we have that
\begin{eqnarray}
\label{exp7}
|| \Lambda_1 \! \otimes  \openone [ | \psi \rangle \langle \psi | ] \! - \!
| \psi \rangle \langle \psi | \, ||_{12}^2 = \! \sum_{k,l=1}^{n_\psi} \beta_k^2 \beta_l^2 \:
{\rm Tr}_1 V_{kl} ( V_{kl})^\dagger \! ,  \; \;
\end{eqnarray}
where $ V_{kl} =  \; \Lambda_1 [ | k \rangle_1 \langle l | ] - | k \rangle_1 \langle l |$.
At this point we apply  the relations Eqs.~(\ref{ape1}), (\ref{ape2}) and (\ref{ape3})
(proved in \ref{appa} and \ref{appb})
and Eq.~(\ref{con1}) that establish the following inequalities
\begin{eqnarray*}
{\rm Tr}_1 \; V_{kl} (V_{kl})^\dagger & \leq & 2 \epsilon^2\; ; \hspace{1.4cm} \forall k \neq l \; ; \\
{\rm Tr}_1 \; V_{kk} (V_{kk})^\dagger & \leq & \epsilon^2\; ;
\hspace{1.6cm} \forall k \;  .
\end{eqnarray*}
These inequalities bound each contribution (trace term) in the sum on the right of Eq.~(\ref{exp7}).
If we replace each term with the corresponding bound and maximize over all possible
$\beta_j$ then we do estimate the expression on the left-hand side of the last equality as
\begin{eqnarray*}
\label{res2}
|| \; \Lambda_1 \otimes \openone [ \; | \psi \rangle \langle \psi | \; ] -
| \psi \rangle \langle \psi | \; ||_{12}^2  & \leq & ( 2 - 1/d ) \, \epsilon^2  ,
\end{eqnarray*}
where $d$ is the dimension of the smaller of the two Hilbert spaces ${\cal H}_1$ and ${\cal H}_2$
in case the two subsystems $1$ and $2$ are different~\footnote{If the two subsystems $1$ and $2$
are different then the number of elements in the Schmidt decomposition Eq.~(\ref{exp12}) $n_\psi$
is always less or equal to $d$ - the dimension of the smaller of the two Hilbert spaces ${\cal H}_1$ and
${\cal H}_2$. Consequently, the number of coefficients $\beta_j$ we maximize over is always bounded
by this number, which in turn bounds the maximum.}.
Since the result is independent of both the map $\Lambda_1$
and the state $| \psi \rangle \langle \psi |$ it holds for all maps $\Lambda_1$ and all
density operators $| \psi \rangle \langle \psi |$ (representing pure states).
Consequently, the supremum over all maps $\Lambda_1$ and all pure states $\rho_{12}$
is less or equal to this value and so is the distance~(\ref{con0})
\begin{equation}
\label{j3}
D_{12} ( \Lambda_1 \otimes \Lambda_2 [ \rho_{12} ] , \rho_{12} ) \leq 2 \sqrt{2 - 1/d} \: \epsilon  .
\end{equation}
The bound (\ref{j3}) is valid for entangled as well as separable states. However,
for separable states we have already found a tighter bound 
$\sqrt{ 2+ 2 \sqrt{(1-1/d_1)(1-1/d_2)}} \epsilon$ [see Eq.~(\ref{ee14})] which 
means that the entangled states can be affected
by independent channels more strongly than  separable states.

\subsection{Detection of entanglement}
\label{detofent}

The difference in the behavior of separable and entangled states resulted into two different bounds.
The bound for entangled states is  weaker and this bound is obeyed by  entangled as well as
separable states. On the other hand the bound for separable states Eq.~(\ref{ee14}) is tighter and
need not be fulfilled by entangled states. Subsequently, any state that violates
the bound (\ref{ee14})  is necessarily entangled and
a direct product of physical channels can be
exploited as a kind of ``entanglement witness''.
 Let us point out that the entanglement witnesses known in the
 literature, Refs.~\cite{horodecki96, terhal00},
are based on a different approach. They are
constructed using positive but not completely positive
maps (that is non-physical maps) acting on one of the two subsystems
and the non-positivity of the final operator  (output) is the indication of entanglement.
On the contrary, in our case, we have a product of two physical maps
so that a physical (completely positive) map is acting on each of the two subsystems
and the difference between an input and the corresponding output is measured. 
In addition there is a potential advantage in this approach.  Not only the question
 whether a state is entangled or separable can be answered. If we relate the distance to entanglement 
then we could answer the question how much entanglement is shared by two quantum systems.

Similarly as in the case of entanglement witnesses, given a pair of maps,
the detection need not be (and in general is not) perfect.
In other words given a pair of channels  only a subset of the
set of all entangled states violates the bound Eq.~(\ref{ee14}) and those are the only states
that are detected as entangled. Naturally, it is desirable to optimize the detection so that
the whole set of entangled states is detected.
There are several things we can do to optimize the detection of entangled states using quantum channels:
\begin{enumerate}
\item Optimal choice of the distances $D_1(.,.)$, $D_2(.,.)$ and $D_{12}(.,.)$
\item Optimal choice of the pair of channels (maps $\Lambda_1$ and $\Lambda_2)$
and subsequent derivation of the bound for separable states  for that particular choice.
\end{enumerate}
It is obvious that both elements influence detection of entanglement. Let us point out
that the choice of maps is not limited to physical channels.
The problem is usually formulated as follows: Given a density matrix of
a bipartite system how strongly the two subsystems are entangled.
That is we have a complete knowledge of the elements
of the density matrix and we are allowed to execute arbitrary operation (function)
on  the matrix to calculate the entanglement.
Such operation can be non-physical and even non-linear. Construction of entanglement witnesses
using a general class of non-physical but linear maps has been investigated in Ref.~\cite{Horodecki02}.
The authors have showed that with the help of linear maps it is possible to distinguish perfectly
the set of entangled states from the set of separable states.
Here we show that this approach could be useful not only for the problem of detection 
but also for the problem of quantifying entanglement.

Let us express the most general bipartite two-qubit state $\rho_{12}$ using the Pauli operators
$\sigma_j$, $j=1,2,3$
\begin{eqnarray*}
\rho_{12} = \frac{1}{4} \left [  \openone + \sum_{j=1}^3 \alpha_j \sigma_j \! \otimes \! \openone
+ \sum_{k=1}^3 \openone \! \otimes \! \sigma_k + \sum_{j,k=1}^3
\gamma_{jk} \sigma_j \! \otimes \! \sigma_k  \right ] \; ,
\end{eqnarray*}
where $\alpha_j$, $\beta_k$ and $\gamma_{jk}$ for $j,k=1,2,3$ are real parameters. Further, 
consider a linear map $\Lambda_{12}$ 
\begin{equation}
\label{det4}
\Lambda_{12}: \rho_{12} \rightarrow \rho_{12} +
\frac{\epsilon}{4} (- \openone + \sum_{j,k=1}^3  \gamma_{jk} \sigma_j \otimes \sigma_k ) \; .
\end{equation}
With the help of the map (\ref{det4}) and the trace distance
 we define the following function
\begin{equation}
{\cal F} (\rho_{12}) \equiv
\frac{1}{\epsilon} {\rm Tr}_{12} | \Lambda_{12} [ \rho_{12} ] - \rho_{12} | - 1 \; .
\end{equation}
The factor $1/\epsilon$ is there to eliminate the dependence on the epsilon
while the $-1$ has been added for a convenience only.
The function ${\cal F}$ has the following properties:
\begin{enumerate}
\item ${\cal F} (\sum_j \lambda_j \rho_{12}^j) \leq \sum_j \lambda_j {\cal F} (\rho_{12}^j)$, convexity.
\item ${\cal F} (U_1 \hspace{-0.05cm} \otimes \hspace{-0.04cm} U_2 \: \rho_{12} \: U^\dagger_1  \hspace{-0.04cm} \otimes   \hspace{-0.03cm} U_2^\dagger) =
{\cal F} (\rho_{12})$,
local unitary equivalence $\forall U_1$ and $\forall U_2$.
\item ${\cal F} (\rho_{12}) \geq 0$, $\forall \rho_{12}$, non-negativity.
\item ${\cal F} (\rho_{12}) = 0$, for all separable states.
\item ${\cal F} (\rho_{12}) = C(\rho_{12})$, where $\rho_{12}$ is pure or Werner state
(for definition of the Werner state see Ref.~\cite{werner89})
and $C(\rho_{12})$ is the concurrence (see Ref.~\cite{wootters97}).
\end{enumerate}
Through the extension of the proposed method to non-physical maps
and a suitable choice of the map acting on the joint state $\rho_{12}$  we have managed to construct
a function that detects entanglement on all Werner states.
Moreover, some of the listed properties of the function $\cal F$ are supposed to
be fulfilled by a function that not only distinguishes
separable and entangled states but performs a harder task - measures entanglement between two
quantum systems. Though, the constructed function
 is not a proper measure of entanglement (there are entangled states for which
${\cal F}$ is zero) a suitable extension  might ``correct'' the function so that
all entangled states are detected.

\section{$N$ channels}

In many physical situations it is less convenient to divide the system under consideration into two
large subsystems than into a large number (say $N$) of  smaller (but equal systems).
Typical example is the envisaged   quantum computer composed of small micro-traps each holding a
single qubit.
In such case individual qubits are spatially separated so that the interaction with environment
can be described by local maps $\Lambda_i$ where the index $i$ labels the qubits (or micro-traps).
These maps can be derived phenomenologically or determined
experimentally so their knowledge can be assumed. Obviously, we want to
keep the influence of the environment as small as possible so each of the
maps would satisfy a condition similar to Eq.~(\ref{con1})
\begin{equation}
\label{con2}
D_i (\Lambda_i [ \rho_i ], \rho_i ) \leq \epsilon \, ;\hspace{0.5cm} \forall i=1,..N \; , \;
\forall \rho_i \in {\cal S} ({\cal H}_i ) \, ,
\end{equation}
where $D_i (.,.)$'s are again metrics (distance functions) and ${\cal S} ({\cal H}_i)$ is the set
of all density operators for each $i=1, \ldots N$.
Using these maps we can find out the state of a particular qubit after interaction with
the environment. However, what is more important  is the final state of the whole
system
\begin{equation}
\label{exp9}
\Lambda_1 \otimes \ldots \otimes \Lambda_N [ \rho_{1..N} ] \, ,
\end{equation}
and, in particular, how much the joint state $\rho_{1..N}$ has changed due to
the interaction with the environment. This change can be characterized by a
distance between the original state
$\rho_{1..N}$ and the output of the product of the individual maps given by Eq.~(\ref{exp9})
\begin{equation}
\label{con3}
D_{1..N} ( \Lambda_1 \otimes \ldots \otimes \Lambda_N [ \rho_{1..N} ], \rho_{1..N} ) \; ,
\end{equation}
where the $D_{1..N}$ is a metric (distance function) defined on the set of all density operators
${\cal S}({\cal H}_{1..N})$ of the joint system $1..N$.
Here we use the same definition of the metric (distance) as before and
define the functions  $D_i(.,.)$ for $j=1,\ldots N$ and
$D_{1..N}(.,.)$ with the help of the norm of the corresponding Hilbert-Schmidt space (for
more details see Sec.~\ref{ssebsta})
\begin{eqnarray}
\label{def4}
D_i (\rho, \sigma) & \equiv & || \rho - \sigma ||_i \; ,\\
\label{def5}
D_{1..N} (\rho, \sigma) & \equiv & || \rho - \sigma ||_{1..N} \; .
\end{eqnarray}
Using these definitions it can be shown that the distance in Eq.~(\ref{con3}) is always less or
equal to $N \sqrt{2 - 1/d} \: \epsilon $ where $d$ is the dimension of the Hilbert space ${\cal H}_i$.

We note that the action of the product of local channels $\Lambda_1 \otimes \ldots \otimes \Lambda_N$
on separable states is such that
$D_{1..N} ( \Lambda_1 \otimes \ldots \otimes \Lambda_N [ \rho_{1..N} ], \rho_{1..N} )\leq N \epsilon$. This means,
that the restriction to the set of separable states leads to the decrease of the bound on $D_{1..N}$ by the
factor  $\sqrt{2 - 1/d}$~\footnote{Here we have used the bound (\ref{exp10}) for separable states that  can be easily extended to a multi-partite case.} .

\vskip 0.5cm

To prove the statement  we will use a very similar line of reasoning as in the case of two subsystems.
First, taking advantage of  the triangle inequality
 we  bound the distance in Eq.~(\ref{con3})
as follows:
\begin{eqnarray}
\label{exp4}
D_{1..N} ( \Lambda_1 \otimes ..\Lambda_N && [\rho_{1..N} ],\rho_{1..N} ) \leq \nonumber \\ 
&& D_{1..N} ( \Lambda_1 \otimes ..\Lambda_N [\rho_{1..N} ],
 \openone \otimes \Lambda_2 \otimes..\Lambda_N [ \rho_{1..N}] ) \nonumber \\
& & \vdots  \nonumber \\
& & +  D_{1..N} ( \openone \otimes ..\openone \otimes \Lambda_N [\rho_{1..N} ], \rho_{1..N} ) \: .
\hspace{2.4cm}
\end{eqnarray}
Each of the $N$ terms on the right side of the last equation can be rewritten as
\begin{equation}
\label{exp3}
D_{1..N} ( \openone \otimes .. \openone \otimes \Lambda_i \otimes \openone \otimes ..
\openone [ \tilde{\rho}_{1..N}^{(i)}] ,  \tilde{\rho}_{1..N}^{(i)} ) \; ,
\end{equation}
where
\begin{eqnarray*}
\tilde{\rho}_{1..N}^{(i)} = \openone \otimes  .. \openone \otimes \Lambda_{i+1} \otimes \Lambda_{i+2}
\otimes .. \Lambda_N [ \rho_{1..N} ] \; ,
\end{eqnarray*}
so it is sufficient to bound the expression~(\ref{exp3}). 
Next, we divide the whole system into two
 parts, an elementary system $i$ and the rest. 
From this point the proof takes the same lines as in the case of two subsystems discussed
 in the Sec.~\ref{entstates}. 
Therefore  we recall the result Eq.~(\ref{j3}) obtained 
there and refer the reader to the Sec.~\ref{entstates}
for more details. The Eq.~(\ref{j3}) states that
\begin{eqnarray*}
D_{1..N} ( \openone \otimes .. \openone \otimes \Lambda_i \otimes \openone \otimes ..
\openone [ \tilde{\rho}_{1..N}^{(i)}] ,  \tilde{\rho}_{1..N}^{(i)} )  \leq \sqrt{2 - 1/d} \; ,
\end{eqnarray*}
where $d$ is the dimension of the $i$-th  elementary subsystem.
Since we have $N$ terms in the expression on the right in Eq.~(\ref{exp4})
the distance (\ref{con3})  is bounded by
\begin{equation}
\label{exp13}
D_{1..N} ( \Lambda_1 \otimes \ldots \otimes \Lambda_N [ \rho_{1..N} ], \rho_{1..N} )
 \leq N \sqrt{2-1/d}  \: \epsilon \; ,
\end{equation}
where $N$ is the number of elementary subsystems each satisfying the condition~(\ref{def4}) and $d$
is the dimension of the Hilbert spaces ${\cal H}_i$ corresponding to the elementary subsystems.
\vskip 0.5cm

\subsection{Example}

To illustrate the character of changes induced by the local maps on the global state of the whole system
 let us consider a simple model of $N$ qubits undergoing a process of decoherence.
That is the Hilbert spaces ${\cal H}_i$ are two-dimensional and the maps $\Lambda_i$ are
chosen to be
\begin{eqnarray}
\Lambda_i : \frac{1}{2} ( \openone + \vec{\alpha} . \vec{ \sigma }  ) \rightarrow 
\frac{1}{2} \left \{ \openone  + \alpha_3 \sigma_3 + (1-k) [ \alpha_1 \sigma_1 + \alpha_2 \sigma_2 ] \right \}
\, ,
\end{eqnarray}
where $k$ is equal
to $k=\sqrt{2} \epsilon$ in order to fulfill the conditions Eq.~(\ref{con2}).
The action of the map $\Lambda_i$ is such that it
preserves the diagonal elements in basis formed by the eigenvectors of $\sigma_3$
while the non-diagonal elements are suppressed.
Such maps describe the process of dephasing, a particular case of decoherence, since
the vanishing of off-diagonal elements results into states that describe
statistical mixtures.

Consider the initial state of the joint system to be the
Greenberger-Horn-Zeiliner (GHZ) state
\begin{eqnarray}
\rho_{1..N} & = & \frac{1}{2} \left \{ |0 \ldots 0 \rangle \langle 0 \ldots 0 |
+ |0 \ldots 0 \rangle \langle 1 \ldots 1 |  \; \; \right . \nonumber \\
& &  \left .  \hspace{0.3cm} + |1 \ldots 1 \rangle \langle 0 \ldots 0|
+ |1 \ldots 1 \rangle \langle 1 \ldots 1 | \right \}  . \;
\end{eqnarray}
 The action of the map $\Lambda_1 \otimes .. \Lambda_N$ on the state $\rho_{1..N}$ described above
can be evaluated straightforwardly and we obtain
\begin{eqnarray}
\Lambda_1 \otimes \ldots  \otimes \Lambda_N [ \rho_{1..N} ] & = & 
 \frac{1}{2} \; \left \{ |0 \ldots 0 \rangle \langle 0 \ldots 0 | +
|1 \ldots 1 \rangle \langle 1 \ldots 1 | \; \; \right . \nonumber \\
&& \left .  + (1-k)^N ( |0 \ldots 0 \rangle \langle 1 \ldots 1 | + h. c. ) \right \} \; . \;
\end{eqnarray}
Despite the fact that the state of each individual qubit remains unchanged (a consequence of this 
is that the conditions (\ref{con2}) are trivially fulfilled)
the state of the whole
system changes because the off-diagonal elements are strongly suppressed. The distance~(\ref{con3})
between the input $\rho_{1..N}$ and the corresponding output
$\Lambda_1 \otimes \ldots \Lambda_N [ \rho_{1..N}]$ gives 
\begin{eqnarray*}
D(\Lambda_1 \otimes .. \otimes \Lambda_N [ \rho_{1..N} ], \rho_{1..N} )  = 
\sqrt{\frac{1}{2}  \left [ 1 + (1-k)^{2N} - 2 (1-k)^N \right] } \; ,
\end{eqnarray*}
which for $\epsilon$ being very small can be estimated as
\begin{equation}
D(\Lambda_1 \otimes .. \otimes \Lambda_N [ \rho_{1..N} ], \rho_{1..N} )  \approx N \epsilon \; .
\end{equation}
The deviation of the GHZ state under the action of the direct product of local maps $\Lambda_i$
for sufficiently small $\epsilon$ scales as $N \epsilon$ which confirms our more general result
Eq.~(\ref{exp13}). Though the result may seem to be optimistic (one might expect worse scaling
with $N$) the effect of the action  of local maps is to disentangle the qubits (destroy quantum
correlations between the qubits).
In addition,  the disentanglement itself is strong since the off-diagonal elements are suppressed 
exponentially with the increase of the number of systems involved in the dynamics.
This example nicely illustrates that though the deviation expressed with the help of the distance
(\ref{con3}) scales as $N \epsilon$ the entanglement
may be destroyed much more dramatically. 

Finally note that in this example the bound for separable states $N \epsilon$, derived 
with the help of Eq,~(\ref{exp10}), is not violated 
in spite of the fact that we have
used an entangled state. We have already pointed out that it is not necessary for any 
entangled state to violate the separable bound. 
To show that the bound can be violated indeed one can choose 
the map $\Lambda_1$ defined in Sec.~\ref{entstates} for maps $\Lambda_i$ and the initial 
state of the form $|bell\rangle^{\otimes N/2}$ where $|bell\rangle$ denotes one of the 
Bell states (see, for instance, Sec.~\ref{entstates}).

\section{Conclusion}
\label{conclusion}

We have analyzed the direct product of linear maps that describe local actions of
 a set of quantum channels.
We have found a bound on the action of such product of maps
(expressed as a distance between an input and output of the product)
provided the linear maps composing the product are bounded as well.
We have addressed two typical scenarios. In the first, a quantum system is
divided into two subsystems and the product
is composed of two maps acting on the two subsystems, respectively.
 In the second scenario a joint system is composed of $N$ equal
 subsystems and we have $N$ linear maps acting on
$N$ subsystems of a given quantum system.

Our analysis has shown that the fundamental difference between the set of
 separable and entangled states yields two different bounds.
 For separable states the distance Eq.~(\ref{con0})
is bounded by $\sqrt{ 2+ 2 \sqrt{(1-1/d_1)(1-1/d_2)}} \epsilon$ 
while in the case of entangled states
 the distance can be larger and is
bounded from above by $2 \sqrt{2 - 1/d} \: \epsilon$.

Let us note that the bound for separable states (\ref{ee14}) is
optimal. That is   
there exists a pair of channels $\Lambda_1$ and $\Lambda_2$ such that the bound is saturated
(examples are presented in \ref{appc}). 
It is interresting to note that the channels that saturate the bound 
are the same channels that saturate the separable bound (\ref{exp10}) for the case of the trace 
distance (see \ref{appc}) or the bound (\ref{j3}) for entangles states 
in case of two-dimensional systems. 
Clearly, to establish the upper bound on the action of a pair of local quantum
channels it is sufficient to find a pair of channels for which the action is maximal 
and set the bound to this maximum. The form of the channels may depend on the
dimensions $d_1$ and $d_2$. However, our results suggest that the channels for
which the bounds are maximal are of the same form for arbitrary $d_1$ and $d_2$ and
are the channels that we have used in our examples.


In the end let us point out that our analysis is not restricted to the
 case of physical maps only and can be extended to the case of linear and hermicity
 preserving maps that are not physical.

\ack
This work was supported in part by the European
Union  projects QGATES, QUPRODIS,  and CONQUEST,  by the Slovak
Academy of Sciences via the project CE-PI, by the project
APVT-99-012304 and by the Alexander von Humboldt Foundation.

\appendix
\section{}
\label{appa}
We prove the relation
\begin{eqnarray}
\label{ape1}
|| \Lambda [ \; | k \rangle \langle l | \; ] - | k \rangle \langle l | \; ||^2  \; = \;
|| \Lambda [ \; | l \rangle \langle k | \; ] - | l \rangle \langle k | \; ||^2  \;  = \; \nonumber \\
\frac{1}{4} \left \{ || \Lambda [ \; ( | k \rangle \langle l | + | l \rangle \langle k |) \; ]
 -  ( | k \rangle \langle l | + | l \rangle \langle k | ) \; ||^2 \; \; \right . \nonumber \\
\left . + \; || \Lambda [ \; ( | k \rangle \langle l | - | l \rangle \langle k |) \; ]
 -  ( | k \rangle \langle l |  - | l \rangle \langle k | ) \; ||^2  \right \} . \; \;
 \end{eqnarray}
for all physical (linear, hermicity preserving and completely positive) maps $\Lambda$
with the norm defined in Eq.(\ref{def1}) and $k \neq l$.

Let us denote by $V_{kl}$ the expression
$ \Lambda [ \, | k \rangle \langle l | \, ] - | k \rangle \langle l | $.
Using the definition of the norm in Eq.~(\ref{def1}) we have that for
any physical map $\Lambda$
\begin{eqnarray*}
|| \Lambda [ \; | k \rangle \langle l | \; ] - | k \rangle \langle l | \; ||^2  & = & {\rm Tr} \;
V_{kl} V_{kl}^\dagger \; ,\\
|| \Lambda [ \; | l \rangle \langle k | \; ] - | l \rangle \langle k | \; ||^2  & = & {\rm Tr} \;
 V_{kl}^\dagger V_{kl} \; ,\\
|| \Lambda [ \; ( | k \rangle \langle l | + | l \rangle \langle k |) \; ]
 -  ( | k \rangle \langle l | + | l \rangle \langle k | ) \; ||^2
& = & {\rm Tr} \; ( V_{kl}+V_{kl}^\dagger)( V_{kl}^\dagger +V_{kl}) \; , \\
|| \Lambda [ \; ( | k \rangle \langle l | - | l \rangle \langle k |) \; ]
 - ( | k \rangle \langle l | -  | l \rangle \langle k | ) \; ||^2
 & = & {\rm Tr} \; (V_{kl} - V_{kl}^\dagger) (V_{kl}^\dagger - V_{kl}) \; .
\end{eqnarray*}
Eq.~(\ref{ape1}) is a direct consequence of the last result.

\section{}
\label{appb}
In this appendix we prove two relations
\begin{eqnarray}
\label{ape2}
|| \; \Lambda [ \; ( | k \rangle \langle l | + | l \rangle \langle k |) \;  ] -
( | k \rangle \langle l | + | l \rangle \langle k | ) \; || &  \leq & 2 \epsilon , \;  \\
\label{ape3}
|| \; \Lambda [ \; ( | k \rangle \langle l | - | l \rangle \langle k |) \;  ] -
( | k \rangle \langle l | - | l \rangle \langle k | ) \; || & \leq & 2 \epsilon , \;
\end{eqnarray}
for all physical (linear, completely positive and hermicity preserving) maps $\Lambda$
satisfying the condition given by Eq.~(\ref{con1}) and $k \neq l$.
The two expressions
\begin{eqnarray*}
&  || \; \Lambda [ \; ( | k \rangle \langle l | + | l \rangle \langle k |) \;  ] -
( | k \rangle \langle l | + | l \rangle \langle k | ) \; || & \; , \\
&  || \; \Lambda [ \; ( | k \rangle \langle l | - | l \rangle \langle k |) \;  ] -
( | k \rangle \langle l | - | l \rangle \langle k | ) \; || & \; ,
\end{eqnarray*}
can be rewritten as
\begin{eqnarray*}
|| \; \Lambda [ \; ( \rho_1 - \rho_2 )\;   ] - (\rho_1 - \rho_2) \; || \; , \\
|| \; \Lambda [ \; i ( \rho_3 - \rho_4 )\;   ] - i (\rho_3 - \rho_4) \; || \; ,
\end{eqnarray*}
where
\begin{eqnarray*}
\rho_1 & = & \frac{1}{2} \; ( | k \rangle + | l \rangle ) ( \; h.c. \; ) \; , \\
\rho_2 & = & \frac{1}{2} \; ( | k \rangle - | l \rangle ) ( \; h.c. \; ) \; , \\
\rho_3 & = & \frac{1}{2} \; ( | k \rangle + i | l \rangle ) ( \; h.c. \; ) \; , \\
\rho_4 & = & \frac{1}{2} \; ( | k \rangle - i | l \rangle ) ( \; h.c. \; ) .
\end{eqnarray*}
By using the triangle inequality 
\begin{eqnarray*}
|| \; \Lambda [ \; ( \rho_1 - \rho_2 )\;   ] - (\rho_1 - \rho_2) \; || & \leq & 
|| \; \Lambda [ \;  \rho_1 \; ] - \rho_1 \; || + || \; \Lambda [ \;  \rho_2 \; ] - \rho_2 \; || \; ,\\
|| \; \Lambda [ \; i ( \rho_3 - \rho_4 )\;   ] - i (\rho_3 - \rho_4) \; || & \leq &  
|| \; \Lambda [ \;  \rho_3 \; ] - \rho_3 \; || + || \; \Lambda [ \;  \rho_4 \; ] - \rho_4 \; || \; ,
\end{eqnarray*}
we obtain the relations Eq.~(\ref{ape2}) and Eq.~(\ref{ape3}) owing to the conditions 
Eq.~(\ref{con1}).

\section{}
\label{appc}
Here we present an example showing that the bounds (\ref{exp10}) and (\ref{ee14}) are optimal.
In this example we will consider a more general case of distances $D_1(.,.)$, $D_2(.,.)$ and 
$D_{12}(.,.)$ and define the distances  with  $p$-norms
\begin{eqnarray*}
D_a^p (\rho_a, \sigma_a) = \left ( {\rm Tr} | \rho_a - \sigma_a |^p \right )^{1/p} \; ,
\end{eqnarray*}
where $a$ labels the system $1$, $2$ or $12$, $\rho_a$ and $\sigma_a$ are density 
operators and $p$ is a positive integer.
The map $\Lambda_1$ is chosen to be a contraction of the form 
\begin{eqnarray*}
\Lambda_1 [ \rho_1 ] = (1-k_1) \; \rho_1 + k_1 \; \frac{1}{d_1} \openone \; ,
\end{eqnarray*}
where $d_1$ is the dimension of the Hilbert space ${\cal H}_1$ and $k_1$ is the
contraction parameter. In what follows we assume 
  $k_1 = \epsilon/ [ (1-1/{d_1})^p + (d_1 - 1)/{d_1^p} ]^{1/p}$
so that the condition (\ref{con1}) is fulfilled.
Similarly, the map $\Lambda_2$ is a contraction
\begin{eqnarray*}
\Lambda_2 [ \rho_2 ] = (1-k_2) \; \rho_2 + k_2 \; \frac{1}{d_2} \openone \; ,
\end{eqnarray*}
where $d_2$ is the dimension of the Hilbert space ${\cal H}_2$ and 
$k_1 = \epsilon/ [ (1-1/{d_2})^p + (d_2-1)/{d_2^p}  ]^{1/p}$.
For the initial state we choose a pure state of the form
 $\rho_{12} = |00 \rangle \langle 00|$. Keeping only terms of the order of $\epsilon$ 
the distance between the input $\rho_{12}$ and the output
 $\Lambda_1 \otimes \Lambda_2 [ \rho_{12} ]$ gives
\begin{eqnarray}
\label{c01}
D^p_{12} ( \Lambda_{1} \otimes \Lambda_2 [ \rho_{12} ] , \rho_{12} ) = \nonumber \\
\left ( \left [ k_1 \left ( 1 \! - \! \frac{1}{d_1} \right ) + k_2 \left ( 1 \! - \! \frac{1}{d_2} \right )
\right ]^p + \left [ \frac{k_1}{d_1} \right ]^p \! (d_1 \! - \! 1) + \left [ \frac{k_2}{d_2} \right ]^p
 \! (d_2 \! - \! 1) \right )^{1/p}  .
\end{eqnarray}
\mbox{} 

\noindent
{\em Case study: Trace distance}

The trace distance is defined with the help of the $1$-norm so that $p=1$. The two contraction 
parameters $k_1$ and $k_2$ read $k_1 = \epsilon / [ 2 (1-1/d_1)]$ and
 $k_2 = \epsilon / [ 2 (1 - 1/d_2)]$. Using these relation in Eq.~(\ref{c01}) the distance between 
the input and the output reads
\begin{eqnarray}
D_{12} (\Lambda_1 \otimes \Lambda_1 [ \rho_{12} ], \rho_{12} ) = 2 \epsilon \; .
\end{eqnarray} \mbox{} 

\noindent
{\em Case study: Hilbert-Schmidt distance}

The Hilbert-Schmidt distance is defined with the help of the $2$-norm so that $p=2$. The two 
contraction parameters $k_1$ and $k_2$ read $k_1 = \epsilon / \sqrt{1-1/d_1}$ and 
$k_2 = \epsilon / \sqrt{1-1/d_2}$. Using these relations in Eq.~(\ref{c01}) the distance between the 
input and the corresponding output reads
\begin{eqnarray}
D_{12} (\Lambda_1 \otimes \Lambda_1 [ \rho_{12} ], \rho_{12} ) = 
\sqrt{2 + 2 \sqrt{(1 - 1/d_1) (1-1/d_2)}} \; \epsilon \; .
\end{eqnarray}



\end{document}